# Research on the impact of asteroid mining on global equity

He Sun[1], Junfeng Zhu[2], Yipeng Xu[3]

University of Nottingham Ningbo China

## Abstract

In the future situation, aiming to seek more resources, human beings decided to march towards the mysterious and bright starry sky, which opened the era of great interstellar exploration. According to the Outer Space Treaty, any exploration of celestial bodies should be aimed at promoting global equality and for the benefit of all nations.

**Therefore,** based on the realistic global situation and the space exploration capabilities of various countries, our goal is to establish a model that can measure the impact of space mining on global fairness, and formulate reasonable policies so that space exploration can ultimately benefit all mankind. We divide this work into 3 stages.

**Firstly,** we defined global equity and set a **Unified Equity Index (UEI)** model to measure it. Then we considered 9 factors related to global equity: Gini coefficient, per capita GDP, employment rate, education index, number of patents, scientific research investment, ecological index, environmental index, and stability index. We merge the factors with greater correlation, and finally, get 6 elements, and then use **the entropy method (TEM)** to find the dispersion of these elements in different countries. Then use **principal component analysis (PCA)** to reduce the dimensionality of the dispersion, and then use the scandalized index to obtain the global equity.

**Secondly**, we simulated a future with asteroid mining and evaluated its impact on **Unified Equity Index (UEI)**. Then, we divided the mineable asteroids into three classes with different mining difficulties and values, identified 28 mining entities including **private companies**, **national** and **international organizations**. We considered changes in the asteroid classes, mining capabilities and mining scales to determine the changes in the value of minerals mined between 2025 and 2085. We convert mining output value into mineral transaction value through allocation matrix. Based on **grey relational analysis (GRA)**, we determined the grey correlation coefficient between the mineral transaction value and the nine factors in a representative country from 2010 to 2019, which represent the effects of the mining industry on the nine different fields. We substitute the factors updated over the years into Model 1 to obtain the changes in global fairness in 60 years.

**Finally, we presented three possible versions of the future of asteroid mining** by changing the conditions. We **propose two sets of corresponding policies** for changes in future trends in global fairness with asteroid mining. We test the separate and combined effects of these policies and find that they are positive, strongly supporting the effectiveness of our model.

[1] scyhs4@nottingham.edu.cn
[2] scyjz14@nottingham.edu.cn
[3] ssyyx20@nottingham.edu.cn

# 1 Introduction

## 1.1 Problem Background

As human society continues to develop, the demand for various minerals is increasing. However, as a non-renewable resource, mineral resources are gradually depleted as they are mined. Due to the fact that the total amount of minerals on the planet is finite, continuous advances in resource recovery technology cannot fundamentally solve the problem of mineral depletion. In this context, the significance of asteroid mining is becoming more apparent. United Nations hopes to develop policies to improve global equity when asteroid mining becomes widespread.

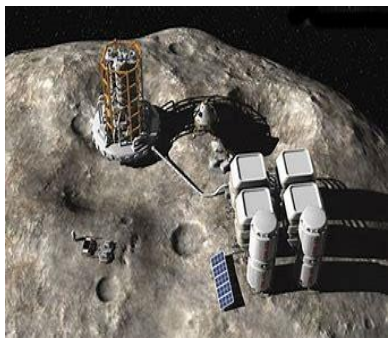 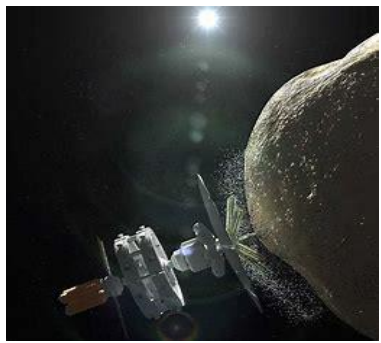 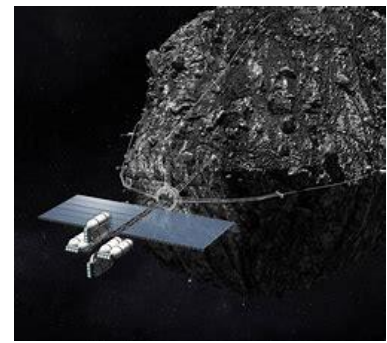

(a) Mining base on an asteroid     (b) Mining airship     (c) Traction airship

**Figure 1: Concept drawings of asteroid mining**

(a) **Mining base on an asteroid**: Build mining bases on asteroids and launch minerals mined back to Earth.
(b) **Mining airship**: Direct mining of asteroids by spacecraft.
(c) **Traction airship**: Tractor asteroids to near-Earth orbit for subsequent mining.

## 1.2 Restatement of the Problem

Considering the background information and restricted conditions identified in the problem statement, we need to solve the following problems:

- Developing an index to measure global equity.
- Predicting the patterns of asteroid mining sector and its impact on global equity.
- Predicting possible changes in the pattern of asteroid mining sector and the impact of these change.
- Amending the Outer Space Treaty to enable the asteroid mining industry to develop while promoting global equity.

## 1.3 Our work

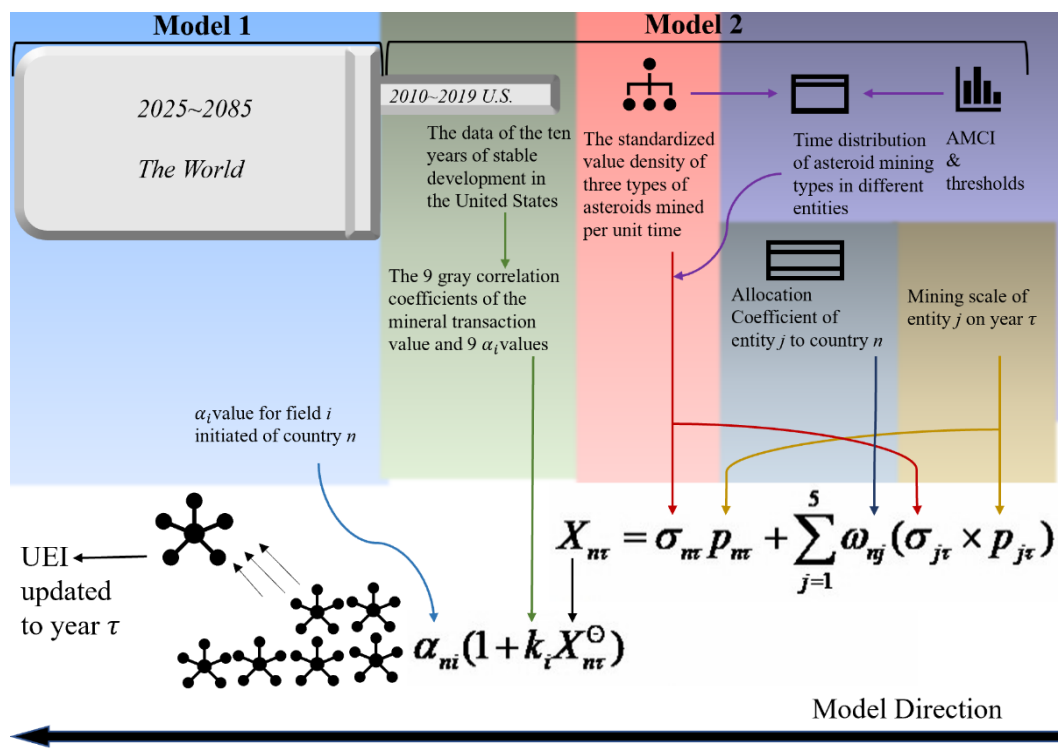

Figure 2: Model Overview

1. We first construct a model that provides an adequate measure of global equity.
2. Next, we develop a complete set of benefit distribution mechanisms for possible changes in the space mining industry.
3. Then, we analyze the relevance of the ore industry to the factors that measure global equity and allocate the benefits gained in the space mining industry to these factors.
4. Finally, we develop sound policies to ensure that the development of the space mining industry enhances global equity and achieves real benefits for all of humanity.

## 2 Assumptions and Justifications

We make the following basic assumptions in order to simplify the problem. Each of our assumptions is justified and consistent with the basic fact.

- **Assume that the correlation between the volume of mineral transactions and these areas is stable, that is, it does not generate accidents that would destroy the relationship between minerals and these areas.**
- ➢ Justification: Since the curve of change in the development of these fields and the curve of change in the volume of mineral transactions in these representative countries we have selected has a certain regularity in the similarity of geometry, and this law has not changed in the last decade or so, we assume that the law is stable for a certain number of years.
- **Assume that minerals from asteroids and minerals from the Earth have the same value in the market.**
- ➢ Justification: The nature of minerals obtained from asteroids is not different from those

obtained on Earth, and their industrial value should be the same.
- **Assume that there is no malicious competition or mutual obstruction in the mining of asteroids, there will be no decline in the mining capacity of a country's asteroids.**
  - Justification: In the first few years of mining, the asteroid resources in space are abundant enough to make more profit than competition by increasing the amount of mining.
- **Assume that the resources of asteroids available for mining will not be depleted in 60 years.**
  - Justification: Scientists estimate that on average, the resources in each asteroid can be mined by humans for about 137 years.[1]
- **Assume that a country with the ability to mine higher-value asteroids will not mine relatively low-value asteroids.**
  - Justification: In the absence of a special purpose, a country should maximize the benefits of mining.
- **Assume that no accident occurs that increases the mining threshold or decreases the mining value of an asteroid (e.g., an asteroid is hit by a meteorite, or an asteroid explodes, etc.).**
  - Justification: This probability is not considered in this model because it is difficult to count.

# 3 Model Preparation

## 3.1 Notations

The key mathematical notations used in this paper are listed in Table 1.

Table 1: Notations used in this paper

| Symbol | Description |
|---|---|
| $UEI$ | Unified Equity Index |
| $T_{eco}$ | Economic Entropy Value |
| $T_{edu}$ | Education Entropy Value |
| $T_{tec}$ | Technology Entropy Value |
| $T_{hea}$ | Health Entropy Value |
| $T_{env}$ | Environment Entropy Value |
| $T_{sta}$ | Stability Entropy Value |
| $\lambda_{neco}$ | Economic Three-dimensional Composition Indicator |
| $\lambda_{ntec}$ | Technology Two-dimensional Composition Indicator |
| $\alpha_{n1}$ | Factor determined by Gini coefficient |
| $\alpha_{n2}$ | Factor determined by GDP per capita |
| $\alpha_{n3}$ | Factor determined by unemployment |
| $\alpha_{n4}$ | Factor determined by educated people |
| $\alpha_{n5}$ | Factor determined by patents |
| $\alpha_{n6}$ | Factor determined by research expenditure |
| $\alpha_{n7}$ | Factor determined by Health Concentration factor |
| $\alpha_{n8}$ | Factor determined by Ecological Indictor |
| $\alpha_{n9}$ | Factor determined by Crime rate |
| $P$ | Percentage of profit from asteroid mining for an entity |

## 3.2 The Data

The data since 2020 are highly variable due to the impact of COVID-19.We collected relevant data from 2010 to 2019 for each country to ensure the long-term validity of the model.

Table 2: Data source collation

| Data names | Database Names |
|---|---|
| Gini coefficient | World Bank |
| GDP per capita | World Bank, IMF |
| Unemployment rate | OECD |
| Education | OECD, EACEA |
| Number of patents | OECD |
| Scientific research investment | OECD |
| Life expectancy class | NCBI, JSTOR |
| Ecological Indicator | Yale University |
| Stability Indicator | NUMBEO |
| Mineral consumption | UNdata |

# 4 Model I: UEI for Measuring Global Equity

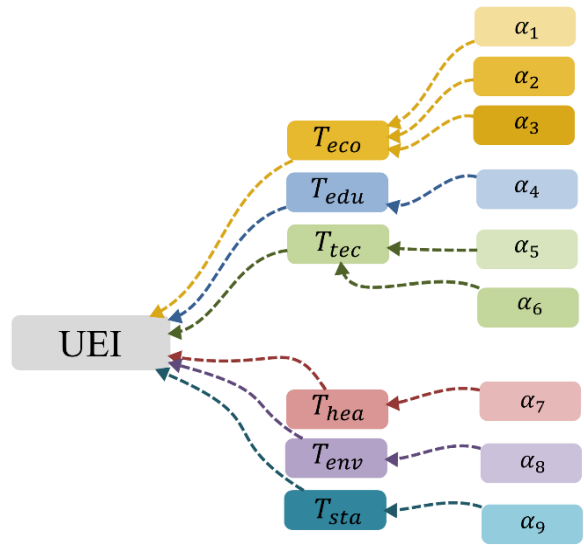

Figure 3: Model I Overview

The definition of global equity is an indicator of whether global resources are distributed fairly to each country. We have designed the Unified Equity Index (UEI) to measure global equity, which depends on the Entropy Values for each aspect. For this purpose, we have selected 23 typical countries based on G20 countries to calculate the UEI. The Entropy Values are de-termined by the factors for each country, using Entropy method to produce the Entropy Values. The six Entropy Values measure global inequity in economic, educational, technological, health environmental and stability, essentially covering the most important aspects of social development. UEI was obtained after principal component analysis and forwarding of Entropy Values.

The steps to calculate the Entropy Values from the factors of each country are as follows

$$P_{ni} = \frac{Z_{ni}}{\sum_{n=1}^{23} Z_{ni}} \tag{1}$$

$$T = -\frac{1}{\ln 23} \cdot \sum_{n=1}^{23} P_{ni} \cdot \ln P_{ni}, T \in [0,1] \tag{2}$$

Where the *i* is the type of an indicator, *n* is the country code.

## 4.1 Calculation of the Six Entropy Values that Form the UEI

Using Principal Component Analysis, UEI was obtained after linear combination of six Entropy Values $T_{eco}$、$T_{edu}$、$T_{tec}$、$T_{hea}$、$T_{env}$ and $T_{sta}$. The Entropy Values.

### 4.1.1 Calculation of Economic Entropy Value

Since there are various factors that determine economic equity, we obtained an Economic Three-dimensional Composition Indicator through three factors determined by the Gini coefficient, GDP per capita and the unemployment rate, which are standardized and represented by the symbols α₁, α₂ and α₃ respectively.

The derivation of *α₁* is as following

$$\alpha_1 = \frac{(1-gini) - Lower\ Threshold}{Upper\ Threshold - Lower\ Threshold} \tag{3}$$

The derivation of *α₂* is as following

$$\alpha_2 = \frac{GDP\ per\ capita - Lower\ Threshold}{Upper\ Threshold - Lower\ Threshold} \tag{4}$$

The derivation of *α₃* is as following

$$\alpha_3 = \frac{\frac{1}{unemployment\ rate} - Lower\ Threshold}{Upper\ Threshold - Lower\ Threshold} \tag{5}$$

The formula for $\lambda_{eco}$ from α₁, α₂ and α₃ is as follows

$$\lambda_{neco} = \sqrt{\alpha_1^2 + \alpha_2^2 + \alpha_3^2} \tag{6}$$

Performing an entropy calculation on $\lambda_{neco}$ gives $T_{eco}$

### 4.1.2 Calculation of Education Entropy Value

$$\alpha_{n4} = \frac{Number\ of\ people\ with\ no\ upper\ secondary\ education}{Number\ of\ people\ with\ post-secondary\ education} \tag{7}$$

Standardizing $\alpha_{n4}$

$$\frac{\alpha_4 - Lower\ Threshold}{Upper\ Threshold - Lower\ Threshold} \tag{8}$$

According to *Education Watch 2001 : renewed hope daunting challenges* [1], education equity can be characterized by the educational composition of the population, specifically access to basic and post-secondary education.

Performing an entropy calculation on $\alpha_{n4}$ gives $T_{edu}$.

### 4.1.3 Calculation of Technology Entropy Value

According to *Economics of Innovation and new technology, 11* [2], the number of patents and research expenditure are core indicators of research

$$\lambda_{ntec} = \sqrt{\alpha_5^2 + \alpha_6^2} \tag{9}$$

Where $\alpha_5$ is the number of patents and $\alpha_6$ is the ratio of research expenditure to GDP.

Standardizing $\lambda_{ntec}$

$$\frac{\lambda_{ntec} - Lower\ Threshold}{Upper\ Threshold - Lower\ Threshold} \tag{10}$$

Performing an entropy calculation on $\lambda_{ntec}$ gives $T_{tec}$.

### 4.1.4 Calculation of Health Entropy Value

$T_{hea}$ is derived as follows

$$COV(X,Y) = \frac{\sum_{i=1}^{n}(X_i - \bar{X}) \cdot (Y_i - \bar{Y})}{n-1} \tag{11}$$

$$CI = \frac{2 \cdot COV(X,Y)}{\bar{Y}} \tag{12}$$

Where $X$ is a serial number representing income level, $Y$ is the life expectancy of the corresponding income level group, $CI$ is the health concentration factor for a country.

Then, the absolute value of CI after forwarding gives $\alpha_{nhea}$

$$\alpha_{n7} = 1 - |CI| \tag{13}$$

Finally, performing an entropy calculation on $\alpha_{n7}$ gives $T_{hea}$.

### 4.1.5 Calculation of Environment Entropy Value

Standardizing $\alpha_{n8}$

$$\frac{\alpha_{n8} - Lower\ Threshold}{Upper\ Threshold - Lower\ Threshold} \tag{14}$$

After performing an entropy calculation, we get $T_{env}$.

### 4.1.6 Calculation of Stability Entropy Value

According to *Global policing*[3], Crime rates can measure an area's Stability equity.

Standardized $\alpha_{n9}$ is given by the following formula

$$\alpha_{n9} = \frac{(1 - crime\ rate) - Lower\ Threshold}{Upper\ Threshold - Lower\ Threshold} \tag{15}$$

After performing an entropy calculation, we get $T_{sta}$.

## 4.2 Results

The calculated factors for each country in 2019 are shown in the figure.

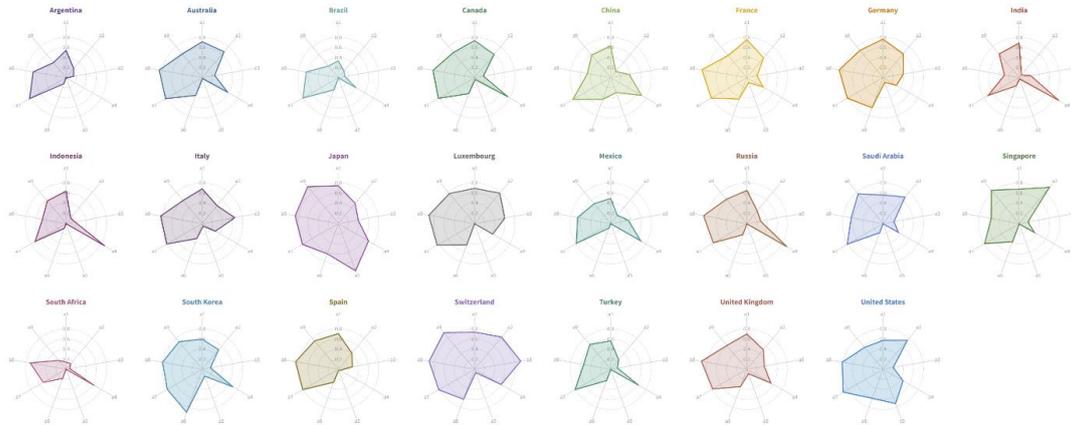

**Figure 4: factors for each country in 2019**

**The final value of the UEI for 2019 was obtained as 0.738146.**

# 5 Model II: Asteroid Mining Model on UEI

According to NASA[4], 2025 is the year of asteroid mining and the market for asteroid mining will reach $30 billion. We have divided asteroids into **three classes according to their mining value: I、II、III**. An **Asteroid Mining Competitive Index** (AMCI) has been defined, which corresponds the value class of an asteroid to the AMCI required to mine it. After determining the AMCI for each entity, we fitted the trend of the AMCI for each entity using a **Logistic Stix curve**. A matrix to represent the class of asteroids that can be mined by each entity

at different time stages. We calculated the impact of the volume of mineral transactions in the United States from 2010 to 2019 on all factors, which determines the UEI, through the Grey Relation Analysis. We use the Grey Relation Coefficient to represent the impact of mineral trading in all countries on the factors that determine social equity.

## 5.1 Ranking the Mining Value of Asteroids

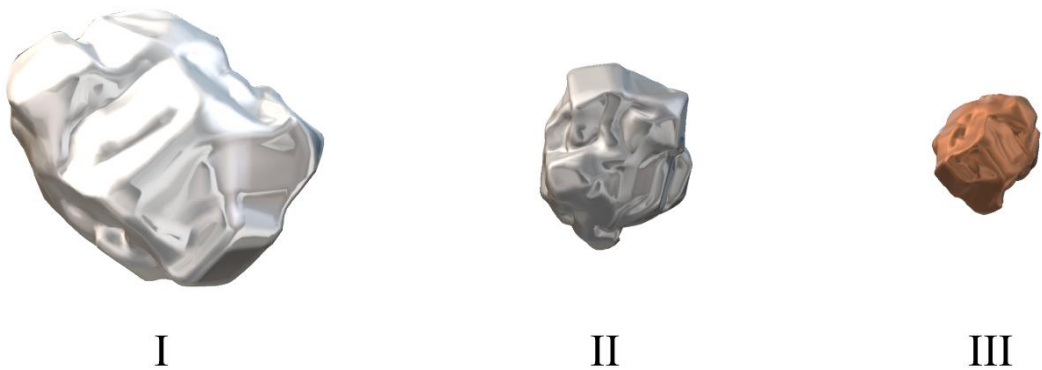

I　　　　　　　　　II　　　　　　　　　III

**Figure 5: Schematic representation of three types of asteroids**

Based on their mining value, the three asteroid classes and their corresponding value are as follows, and then divided the capacity threshold required to mine each of the three asteroids, $h$. Asteroids classified by mining value and their properties are shown in Table 4.

**Table 4: Three Classes of Asteroids**

| Asteroid Classes | I | II | III |
|---|---|---|---|
| h | 37.53 | 32.81 | 22.68 |
| σ | 7.81 | 3.64 | 1.52 |
| Standardized distance to Earth | 234.09 | 198.63 | 110.29 |
| Total price of most valuable minerals (trillions) | 26.99 | 5.73 | 1.66 |
| Elemental symbols for the high value minerals contained | Mn Fe Co Ni Cu Zn Ga Ge Ag Au Mo Ru Rh Pd Ba Re Ir Pb | Mn Fe Co Ni Cu Zn Ga Ge Ag Au Mo Ru | Mn Fe Co Ni Cu Zn Ga Ag |

## 5.2 Entities Mining Asteroids

Based on the Space Competitive Indicator accessed, we have derived the AMCI for each entity according to the expert scores provided by *Aerospace strategy for the Aerospace Nation*[4]. An entity able to mine an asteroid only after the entity's AMCI reaches $h$ for that class of asteroid. Since the highest value class asteroids offer the best benefits when mined, we assume that entities will only mine the highest value class asteroids they can. Because the AMCI of each entity is unpredictable over the next few decades, we use the decline in $h$ to represent the relative rise in the AMCI of each entity. The downward trend of h for the three asteroids is similar to the change in the value of the logistic Stix curve. We use $\tau$ to represent the year starting in 2025 and ending in 2085.

$h_{i\tau}$ is given by the following formula

$$h_{i\tau} = h_i \times \frac{2}{1+e^{0.034\tau}} \quad (16)$$

The calculated AMCI for each entity in 2025 is shown below.

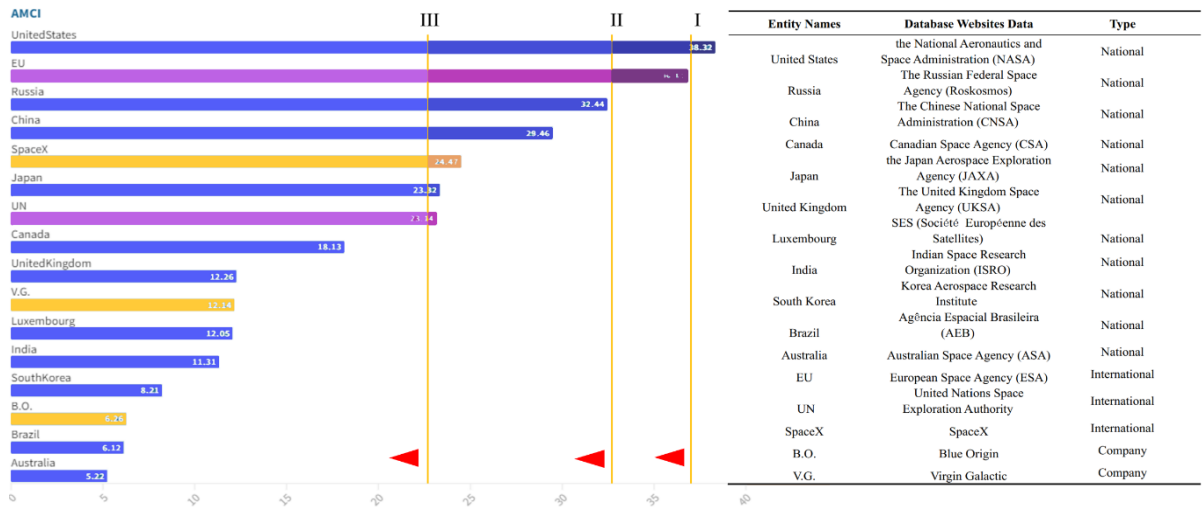

**Figure : AMCI for each entity in 2025**

## 5.3 Matrix Representing the Change

Matrix representing the change in value class of asteroids that an entity can mine over time. The rows of this matrix represent different values of $\tau$ and the columns represent entities.

## 5.4 Determining the Impact

By conducting a Grey Relationship Analysis between the value of $X_{i\tau}^{\odot}$ and factors that determine equity in the United States from 2010 to 2019. We calculated the Grey Relation Coefficient, $k_{ni}$, between $X_{ni\tau}^{\odot}$ and the factors that determine the UEI. The heat map represening the $k_{ni}$ is as follows

$$\alpha_{ni\tau} = \alpha_{ni}(1 + k_i X_{ni\tau}^{\odot}) \qquad (17)$$

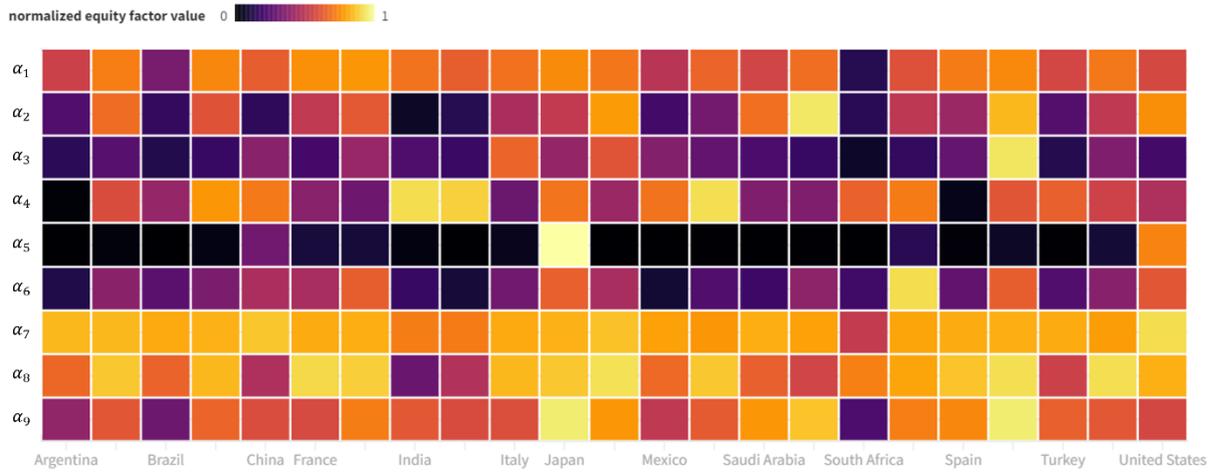

Figure: $k_{ni}$, between $X_{ni\tau}^{\odot}$ and the factors

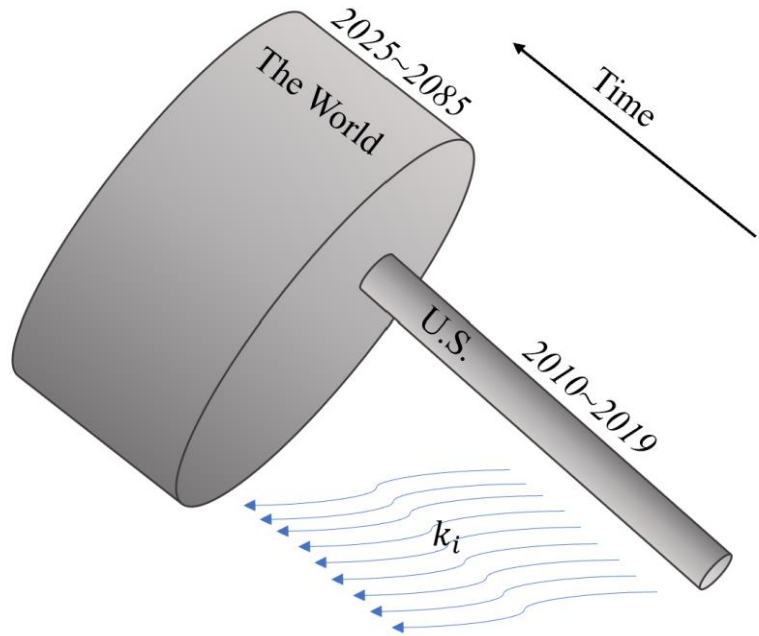

Based on **grey relational analysis (GRA)**, we determined the grey correlation coefficient between the mineral transaction value and the nine factors in a representative country from 2010 to 2019, which represent the effects of the mining industry on the nine different fields. We substitute the factors updated over the years into Model 1 to obtain the changes in global fairness in 60 years.

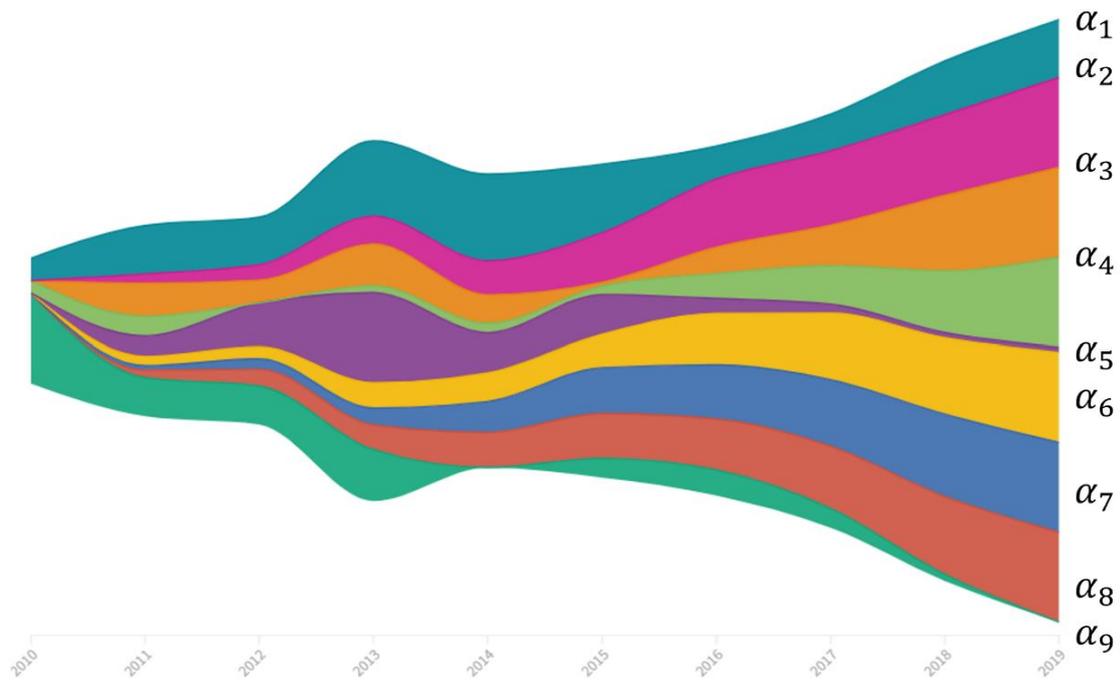

| | EU | UN | SpaceX | V.G. | B.O. | Self-Naion |
|---|---|---|---|---|---|---|
| Argentina | 0 | 0.0434 | 0.0062982 | 0.0062982 | 0.0062982 | 0 |
| Australia | 0 | 0.0434 | 0.0197542 | 0.0197542 | 0.0197542 | 1 |
| B.O. | 0 | 0 | 0 | 0 | 0 | 0 |
| Brazil | 0 | 0.0434 | 0.0265557 | 0.0265557 | 0.0265557 | 1 |
| Canada | 0 | 0.0434 | 0.0246326 | 0.0246326 | 0.0246326 | 1 |
| China | 0 | 0.0434 | 0.2019254 | 0.2019254 | 0.2019254 | 1 |
| EU | 0 | 0 | 0 | 0 | 0 | 0 |
| France | 0.285 | 0.0434 | 0.0384055 | 0.0384055 | 0.0384055 | 0 |
| Germany | 0.283 | 0.0434 | 0.0545962 | 0.0545962 | 0.0545962 | 0 |
| India | 0 | 0.0434 | 0.0405972 | 0.0405972 | 0.0405972 | 0 |
| Indonesia | 0 | 0.0434 | 0.0158231 | 0.0158231 | 0.0158231 | 0 |
| Italy | 0.214 | 0.0434 | 0.0283516 | 0.0283516 | 0.0283516 | 0 |
| Japan | 0 | 0.0434 | 0.0716212 | 0.0716212 | 0.0716212 | 1 |
| Luxembourg | 0.071 | 0.0434 | 0.0010054 | 0.0010054 | 0.0010054 | 1 |
| Mexico | 0 | 0.0434 | 0.0179442 | 0.0179442 | 0.0179442 | 0 |
| Russia | 0 | 0.0434 | 0.0238549 | 0.0238549 | 0.0238549 | 1 |
| SaudiArabia | 0 | 0.0434 | 0.0112134 | 0.0112134 | 0.0112134 | |
| Singapore | 0 | 0.0434 | 0.0052942 | 0.0052942 | 0.0052942 | 0 |
| SouthAfrica | 0 | 0.0434 | 0.0049689 | 0.0049689 | 0.0049689 | 0 |
| SouthKorea | 0 | 0.0434 | 0.0232893 | 0.0232893 | 0.0232893 | 1 |
| SpaceX | 0 | 0 | 0 | 0 | 0 | 0 |
| Spain | 0.143 | 0.0434 | 0.0196976 | 0.0196976 | 0.0196976 | 0 |
| Switzerland | 0 | 0.0434 | 0.0103437 | 0.0103437 | 0.0103437 | 0 |
| Turkey | 0 | 0.0434 | 0.0107665 | 0.0107665 | 0.0107665 | 0 |
| UN | 0 | 0 | 0 | 0 | 0 | 0 |
| UnitedKingdom | 0 | 0.0434 | 0.0400316 | 0.0400316 | 0.0400316 | 1 |
| UnitedStates | 0 | 0.0434 | 0.3030294 | 0.3030294 | 0.3030294 | 1 |
| V.G. | 0 | 0 | 0 | 0 | 0 | 0 |

## 5.5 Estimation of $X_{n\tau}$

We introduced $P_{j\tau}$ as **Mining Scale** of entity *j* on year $\tau$. Its initial value is based on the investment fund estimates published by JP Morgan from the companies, national space agencies and international organizations that we have considered. Then we standardized the source values so that $P_{j\tau}$ is in the range 0~2.

The value of $P_{j\tau}$ conforms to a logistic-like function below.

$$P_{j\tau} = P_{j\tau} \times \frac{2}{1+e^{-0.022\tau}} \tag{18}$$

We introduced $\omega_{nj}$ as **Allocation Coefficient** to transform the total mining value from entity j to the Mineral transaction value for country *n*. This transformation can be thought of as a sale, distribution, or other form of economic phenomena that takes place. We normalize the source values so that $\omega_{nj}$ is in the range 0~1.

For entity *j* that is a private company (SpaceX, Blue Origin, and Virgin Galactic), we set $\omega_{nj}$ to be the normalized proportion of GDP purchasing power parity of country *n*.

For entity *j* that is the UN, we set each $\omega_{nj}$ to be the reciprocal of *n*.

For entity *j* that is the EU, we set each $\omega_{nj}$ to be the normalized value of resource allocation ratios in the European Coal and Steel Community.

For countries with mineral mining capacity, we assume that they only allocate minerals to themselves, so we add $\sigma_m P_m$ to the above formula.

Therefore, **Updated Mineral Transaction Value** $X_{n\tau}$ in each five year from 2025~2085 can be calculated by the following equation.

$$X_{n\tau} = \sigma_{n\tau} p_{n\tau} + \sum_{j=1}^{5} \omega_{nj}(\sigma_{j\tau} \times p_{j\tau}) \tag{19}$$

We simulated a future with asteroid mining and evaluated its impact on Unified Equity Index (UEI). Then, we divided the mineable asteroids into three classes with different mining difficulties and values, identified 28 mining entities including private companies, national and international organizations. We considered changes in the asteroid classes, mining capabilities and mining scales to determine the changes in the value of minerals mined between 2025 and 2085. We convert mining output value into mineral transaction value through allocation matrix.

## 5.6 Results

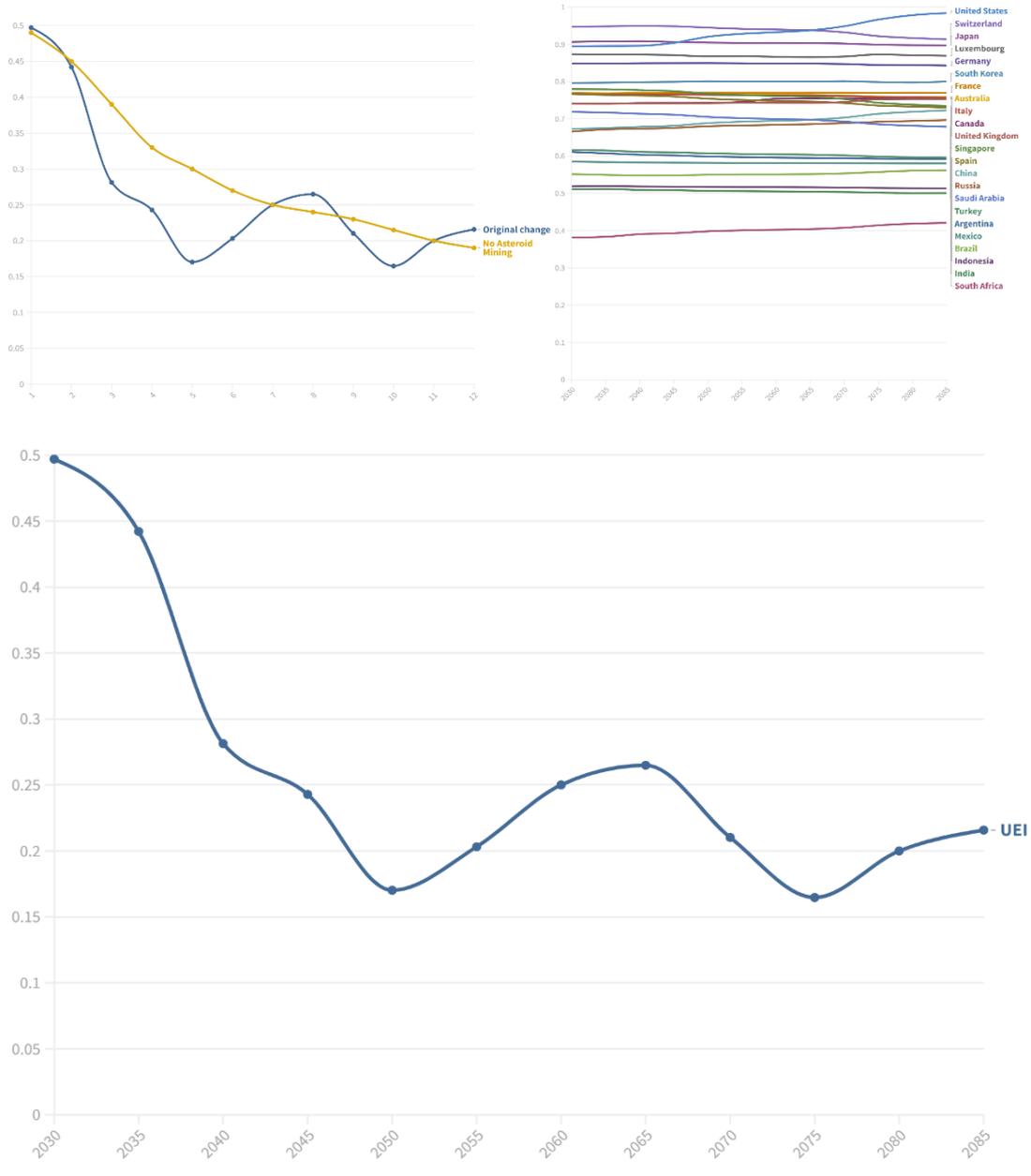

The calculations show that as capital enterprises gain huge benefits from asteroid mining, their capital becomes even larger and the huge profits tempt them to keep inflating their size, thus leading to a disorderly expansion of enterprises. Moreover, due to the profit-seeking nature of capital, companies are more inclined to trade the minerals they acquire with countries that have relatively high indices of mineral purchasing power, and therefore these countries will grow relatively fast, which are often otherwise powerful countries. This will create a further imbalance in global equity.

# 6 Possible changes and impacts

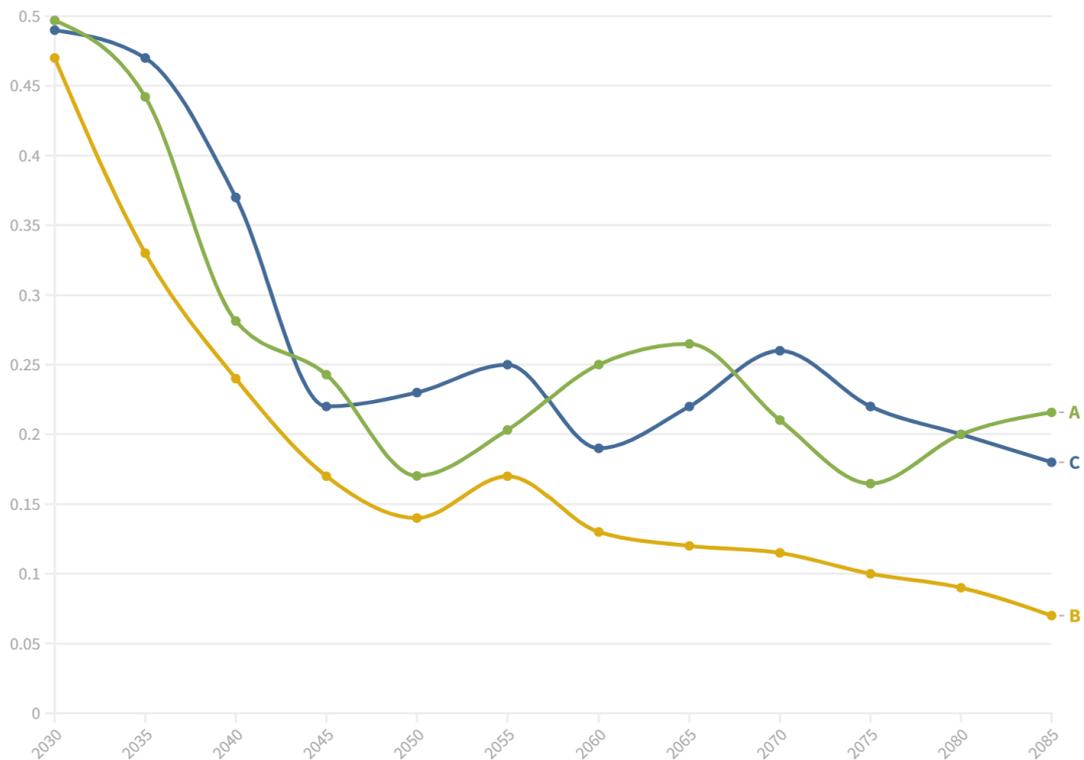

## 6.1 Possible Version A

Since capital enterprises have obtained huge benefits from asteroid mining, the capital of enterprises has become even larger, and the huge profits tempt enterprises to continuously exaggerate their scale, thus resulting in disorderly expansion of enterprises. In addition, due to the profit-seeking nature of capital, companies are more inclined to trade the acquired minerals with countries with relatively high mineral purchasing power index, so these countries will develop relatively faster, and these countries are often originally stronger countries. It will cause further imbalances in global equity.

## 6.2 Possible Version B

Due to the rapid development of the EU in the field of space mining, the relative acquisition of minerals in the EU will be more than other entities in the early stage, and some countries in the EU originally did not have the ability to mine asteroids, but they have obtained the European Coal and Steel Community The allocated space minerals allow more European countries to profit from asteroid mining. In the middle and late stages of asteroid mining, the United Nations is also making great strides in space mining, trading ore to more countries and making more countries profit from it. Therefore, in the vicinity of the nodes where the space mining capabilities of these international organizations have improved, the global fairness has increased to a certain extent.

### 6.3 Possible Version C

Some countries that originally had weak or no space mining capabilities have increased investment in scientific research and developed their own space mining fields while allocating or importing a large amount of minerals, so that all countries can benefit from space mining. profit. However, due to the large gap in the extraction capacity of countries, global equity is still on an overall downward trend. This is one of the most likely scenarios.

## 7 Test and Sensitivity Analysis

We construct a model that represents the impact of space mining on global equity in each country. Our model is based on reasonable space mining assumptions and the characteristics that mining will have an impact on each domain, and objectively predicts changes in the entropy of each domain and changes in the difficulty of mining each type of asteroid. Since we do not have a large dependence on certain data in the process of building our model, our model can still produce possible results corresponding to reality when some of the data change. Whether we add some data or remove some data in our model. The results of the model do not fluctuate much, indicating that our model has relatively good stability and weak sensitivity. We have used principal component analysis (PCA) and entropy method (TEM) to derive the results of the global equity model, so our model has a strong objectivity. After computer testing, our model is not significantly different from the results obtained based on logical inference when we adjust the parameters that change due to policy formulation. Based on the above stability and sensitivity analysis, our model is stable.

## 8 Policies

◆ Policy 1:
Mining Information Policy

After mining an asteroid, an entity should notify the international community of the entity's mining progress on the asteroid, aiming to facilitate the international community to estimate the remaining value of the asteroid and prevent future entities that want to mine the asteroid to discover minerals in the future has been depleted.

◆ Policy 2:
Mineral Legacy Policy

When the remaining resources of the planet are less than 30%, countries with strong space competitiveness should not continue to mine the asteroid, and when the remaining resources of the planet are less than 15%, countries with medium space competitiveness should not continue to mine the asteroid.

The reason for the formulation and the explanation of the reason for the curve change:

These policies are designed to provide protection to less space-competitive entities from becoming unable to operate mining after developing space mining technology in later years. The rise in the middle of the curve is due to a group of countries that have developed the ability to mine space. The level of minerals has risen at the end because another group of originally weaker countries has also developed to the extent that they can mine space minerals

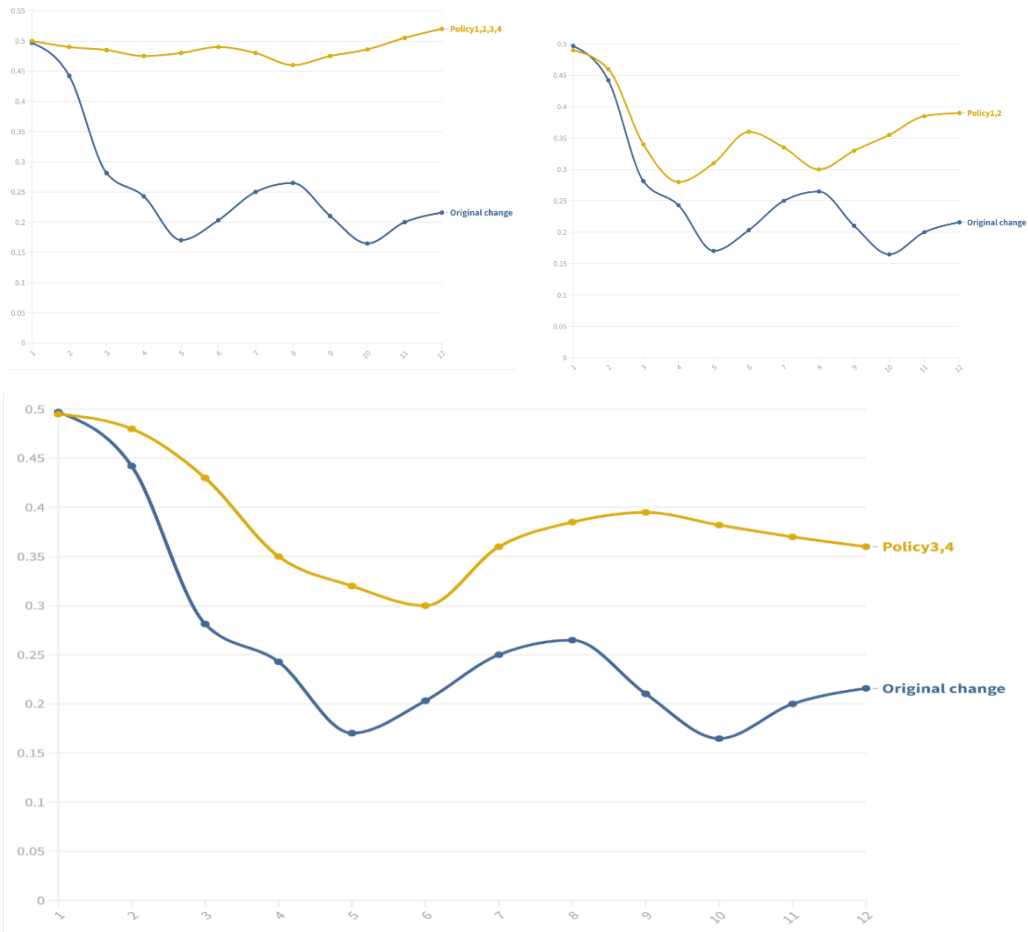

◆ Policy 3:

Mutual Assistance Policy

The policy promotes mutual assistance between countries, encouraging countries with high mining capacity to be allocated to countries with low and no mining capacity. This reflects the purpose of the UN for the benefit of all nations.

◆ Policy 4:

Antitrust Policy

The price of minerals will be subject to certain restrictions to prohibit companies with high mining capacity from abusing their dominant market position.

◆ Policy 5:

Transaction Guidance Policy

The policy guides mining companies to conduct more mining transactions in economically backward countries.

◆ Policy 6:

All the above policies

After the policy is fully implemented, the global fairness decreases to a lesser extent due to space mining, and later the curve rebounds due to the distribution of international organizations and the reason why originally weaker countries have mined ore.

This rise in international equity reflects the philosophy of "all for one, and one for all".

## 9 Strengths and Weaknesses

### 9.1 Strengths

- We use scientific methods to repair the missing part of the data, such as interpolation and fitting methods. Statistical methods such as principal component analysis (PCA), grey correlation analysis and hierarchical analysis (AHP) are also used to approximate our model parameters.
- Our modeling of global equity dynamically takes into account the impact of each domain on global equity.
- Our asteroid mining model takes into account various factors, such as the combination of asteroid orbital distance and asteroid mineral types and values, differences in the competitiveness of asteroid mining among different entities, the relationship between mineral mining and trade volume allocation, changes in the scale of mining among different entities over time, and so on.
- We build our model on the basis of reliable mathematical theory and link it to the characteristics of the change of ore mining difficulty in reality, such as setting the threshold of mining difficulty similar to the change of Logistic Equation and the change of scale of asteroid mining by major entities.
- We have developed and validated the results of policies that are justified at different levels and perspectives.

### 9.2 Weaknesses

- Our model only considers possible impacts 60 years in the future, and our model may not be accurate beyond that time.
- Our model does not consider the impact of the resource curse effect (slow national development due to resource abundance) across countries and sectors.